\magnification\magstep1
\font\cst=cmr10 scaled \magstep3

\vglue 0.5cm
\centerline{\cst Mass and angular momenta}
\vskip 0.5cm
\centerline{\cst of Kerr anti-de Sitter spacetimes in}
\vskip 0.5cm
\centerline{\cst Einstein-Gauss-Bonnet theory} 

\vskip 1.5 true cm
\centerline{\bf Nathalie Deruelle$^*$ {\rm and} Yoshiyuki Morisawa$^{**}$}
\vskip 0.5cm
\centerline{\it $^*$Institut d'Astrophysique de Paris,}
\centerline{\it GReCO, FRE 2435 du CNRS,}
\centerline{\it 98 bis boulevard  Arago, 75014, Paris, France}
\centerline{and}
\centerline{\it Institut des Hautes Etudes Scientifiques,}
\centerline{\it 35 Route de Chartres, 91440, Bures-sur-Yvette,
France}

\bigskip
\centerline{\it$^{**}$  Yukawa Institute for Theoretical Physics}
\centerline{\it Kyoto University}
\centerline{\it Kyoto 606-8502, Japan}\footnote{}{deruelle@ihes.fr

 morisawa@yukawa.kyoto-u.ac.jp}

\medskip

\vskip 0.8cm

\centerline{29 November 2004}

\vskip 1cm
\noindent {\bf Abstract}
\bigskip  

We compute the mass and angular momenta of rotating anti-de Sitter spacetimes in
Einstein-Gauss-Bonnet theory of gravity using a 
superpotential derived from standard N\oe ther
identities. The calculation relies on the 
fact that the Einstein and Einstein-Gauss-Bonnet vacuum equations are the same when linearized
on maximally symmetric backgrounds and uses the recently discovered
D-dimensional Kerr-anti-de Sitter solutions to Einstein's equations.

\vfill\eject

In [1] Gibbons, Lu, Page and Pope found D-dimensional Kerr-anti-de
Sitter solutions of Einstein's equations in vacuo. In [2] Gibbons,
Perry and Pope calculated the angular momenta $J_{(D)i}$ of these rotating spacetimes using Komar's
integrals. They also obtained the mass $E_{(D)}$ of rotating {\it black holes} using the first law of
thermodynamics. 
In [3], the mass of these rotating spacetimes
 was
computed using the
covariant Katz-Bi\v c\' ak-Lynden Bell (KBL) superpotential [4], that is by means of classical N\oe
ther identities.
The results confirm those of [2] and hold 
whether the source is a black hole or a
``star'' because they do not depend on the
actual source of curvature.\footnote{$^{(1)}$}{In the second version of their paper, Gibbons
Perry and Pope showed that the mass can also be obtained using the classical
Ashtekar-Magnon-Das conformal boundary formula.} 

In Einstein-Gauss-Bonnet (EGB) theory in 5 dimensions, the static spherically symmetric
Schwarz\-schild-like solution was obtained in [5] and its mass was
computed by various methods, including the use of the KBL superpotential
extended to the Einstein-Gauss-Bonnet Lagrangian, see [6].

The rotating Kerr-like solution of the EGB equations is not known (apart from the topologically special
solution given in [7]).
However, on one hand, the Einstein and Einstein-Gauss-Bonnet vacuum equations
are the same when linearized on a maximally symmetric background (see
[5] and below) : hence the D-dimensional Kerr-anti-de Sitter solutions
to the Einstein field equations
found in [1] are also asymptotic solutions to the Einstein-{\it Gauss-Bonnet} equations.
On the second hand, only the asymptotic form of the metric is required
to define and compute the mass and angular momenta of a given
spacetime when using superpotentials derived from N\oe ther
identities.

The object of this Note
is to use these two remarks to compute the mass and
angular momenta of Kerr-AdS spacetimes in EGB theory by means of the
KBL superpotential proposed in [6].

\vskip 0.8cm

The Einstein-Gauss-Bonnet Lagrangian can be written as (see e.g. [6])
$$\eqalign{L&\equiv
-2\Lambda+R+\alpha\, R^{\mu\nu\rho\sigma}P_{\mu\nu\rho\sigma}\cr
\hbox{with}\quad
P_{\mu\nu\rho\sigma}&\equiv
R_{\mu\nu\rho\sigma}-2(R_{\mu[\rho}g_{\sigma]\nu}
-R_{\nu[\rho}g_{\sigma]\mu})+R \, g_{\mu[\rho}g_{\sigma]\nu}\cr}\eqno(1)$$
where $\Lambda$ is a (negative) cosmological constant, $\alpha$  a
coupling constant which is zero in Einstein's theory, and
$R_{\mu\nu\rho\sigma}$, $R_{\mu\nu}$ and $R$ the Riemann tensor, Ricci
tensor and curvature scalar of the metric $g_{\mu\nu}$. Brackets stand
for anti-symmetrization : $f_{[\mu\nu]}\equiv {1\over2}(f_{\mu\nu}-f_{\nu\mu})$.
The equations of motion derived from that Lagrangian are
$$R^\mu_\nu+2\alpha\, R^{\mu\lambda\rho\sigma}P_{\nu\lambda
\rho\sigma}-{1\over2}\delta^\mu_\nu\, L=0\,.\eqno(2)$$

We linearize them, that is, we set
$$g_{\mu\nu}=\overline{g}_{\mu\nu}+h_{\mu\nu}+{\cal O}(h^2)\,.\eqno(3)$$
We choose the background $\overline{g}_{\mu\nu}$ to be the metric of 
a maximally symmetric spacetime  $\overline{\cal M}$ solving the field equations
(2). Since they are quadratic in the Riemann tensor there are 
two solutions,
defined by 
$$\overline{R}_{\mu\nu\rho\sigma}=-{1\over{\cal L}^2}(\overline{g}_{\mu\rho}\overline{g}_{\nu\sigma}-\overline{g}_{\mu\sigma}\overline{g}_{\nu\rho})
\quad\hbox{where}\quad {1\over{\cal
L}^2}\equiv{1\over2\tilde\alpha}\left(1\pm\sqrt{1-{4\tilde\alpha\over l^2}}\right)
\eqno(4)$$
with $\tilde\alpha\equiv(D-3)(D-4)\alpha$ and
$l^2\equiv-{(D-1)(D-2)\over2\Lambda}\ $, $D$ being the dimension of
spacetime. We shall choose the lower sign for ${\cal L}^2$ so that the
limit when $\alpha=0$ solves
the Einstein equations. (When $\Lambda$ and $\alpha$ are such that
$\sqrt{1-4\tilde\alpha/ l^2}=0$ the two roots coalesce into a single
double root.) 
It is  an exercise to show that at linear order the field
equations (2) read (a result already obtained in [5] for
$\Lambda=0$)
$$
\sqrt{1-{4\tilde\alpha\over
l^2}}\left[ R^\mu_\nu-{1\over2}\delta^\mu_\nu\, R+\delta^\mu_\nu\,\Lambda
{l^2\over{\cal L}^2}\right]={\cal O}(h^2)\eqno(5)$$
where ($\overline{\nabla}_\mu$ being the covariant derivative
associated to $\overline{g}_{\mu\nu}$)
$$R^\mu_\nu=-{(D-1)\over{\cal
L}^2}(\delta^\mu_\nu-h^\mu_\nu)+ {1\over2}\left[\overline{\nabla}_\alpha(\overline{\nabla}_\nu
h^{\alpha\mu}+\overline{\nabla}^\mu
h_\nu^\alpha-\overline{\nabla}^\alpha h^\mu_\nu)
-\overline{\nabla}^\mu\overline{\nabla}_\nu
h^\rho_\rho\right]+{\cal O}(h^2)\eqno(6)$$
is the linearized Ricci tensor on $\overline{\cal M}$. Hence the
linearized Einstein-Gauss-Bonnet equations are
 the same as the linearized pure Einstein equations with effective
cosmological constant $\Lambda_e\equiv\Lambda{l^2\over{\cal L}^2}$,
and up to the overall factor $\sqrt{1-4\tilde\alpha/l^2}$ that we
shall assume not to be zero.

The general D-dimensional Kerr-AdS metrics solutions to Einstein's
equations found in [1] therefore also solve the EGB equations at linear
order. As an example,
the 5-dimensional Einstein-Kerr-AdS metric reads, in Kerr-Schild
ellipsoidal coordinates
$x^\mu=(t,r,\theta,\phi,\psi)$ [1] :
$$ds^2=d\overline{s}^2+{2m\over U}(h_\mu dx^\mu)^2\eqno(7)$$
with $d\overline{s}^2$ the AdS line element of $\overline{\cal M}$ :
$$\eqalign{d\overline{s}^2=-{(1+r^2/{\cal
L}^2)\Delta_\theta\over\Xi_a\Xi_b}&dt^2+
{r^2\rho^2\over(1+r^2/{\cal
L}^2)(r^2+a^2)(r^2+b^2)}dr^2+{\rho^2\over\Delta_\theta}d\theta^2+\cr
&+{r^2+a^2\over\Xi_a}\sin^2\theta\,
d\phi^2+{r^2+b^2\over\Xi_b}\cos^2\theta\,
d\psi^2\cr}\eqno(8)$$
where $\Delta_\theta\equiv \Xi_a\cos^2\theta+\Xi_b\sin^2\theta$, where
$\rho^2\equiv r^2+ {\cal L}^2(1-\Delta_\theta)$, and where $\Xi_a$ and
$\Xi_b$ are related to the rotation parameters $a$ and $b$ by
$\Xi_a\equiv 1-a^2/{\cal L}^2$, $\Xi_b\equiv 1-b^2/{\cal L}^2$. As for
the function $U$ and the null vector
$h_\mu$ they are given by $U=\rho^2$ and
$$h_\mu
dx^\mu={\Delta_\theta\over\Xi_a\Xi_b}\,dt+{r^2\rho^2\over(1+r^2/{\cal
L}^2)(r^2+a^2)(r^2+b^2)}\,dr+{a\sin^2\theta\over\Xi_a}d\phi+{b\cos^2\theta\over\Xi_b}d\psi\,.\eqno(9)$$
Note that the full Ricci tensor of the Kerr-Schild metric (7-9) is
linear in $h_{\mu\nu}={2m\over U} h_\mu h_\nu$ and hence 
exactly given by (6) [1]. Note also [1] that the metric (7-9) tends for
large $r$ to the AdS metric in static (non rotating) coordinates.

\vskip 0.8cm
 
There are various ways to associate conserved charges to the EGB Lagrangian (see e.g. [6] for a short review). 
We shall use here the superpotential $J^{[\mu\nu]}$ proposed in [6] which is defined
as
$$\hat J^{[\mu\nu]}\equiv\hat J^{[\mu\nu]}_E+\alpha\, \hat
J^{[\mu\nu]}_{GB}\eqno(10)$$
where a hat means multiplication by $\sqrt{-g}$, with $g=\overline {g}$ the
determinant of the metric (7-9). The Einstein contribution  $\hat
J^{[\mu\nu]}_E$ is given by, see [4]
$$-8\pi \hat J^{[\mu\nu]}_E\equiv
D^{[\mu}\hat\xi^{\nu]}-\overline{D^{[\mu}\hat\xi^{\nu]}}+\hat\xi^{[\mu}k_E^{\nu]}$$
$$\hbox{with}\quad
k^\nu_E\equiv
g^{\nu\rho}\Delta^\sigma_{\rho\sigma}-g^{\rho\sigma}\Delta^\nu_{\rho\sigma}\quad\hbox{and}\quad\Delta^\mu_{\nu\rho}\equiv\Gamma^\mu_{\nu\rho}-\overline{\Gamma^\mu_{\nu\rho}}\eqno(11)$$
where $\Gamma^\mu_{\nu\rho}$ are the Christoffel symbols of metric
(7-9) and where an overline means that the quantity is evaluated on the
AdS spacetime $\overline{{\cal M}}$ (that is, by setting $m=0$ in
(7)). 
The following proposal for the Gauss-Bonnet contribution $\hat
J^{[\mu\nu]}_{GB}$ was put forward in
[6] :
$$-8\pi \hat
J^{[\mu\nu]}_{GB}\equiv2\left[P^{\mu\nu\alpha\beta}D_{[\alpha}\hat\xi_{\beta]}-\overline{P^{\mu\nu\alpha\beta}D_{[\alpha}\hat\xi_{\beta]}}\right]
+\hat\xi^{[\mu}k_{GB}^{\nu]}\eqno(12)$$
$$\hbox{with}\quad k^\nu_{GB}\equiv4\left(P^{\nu\alpha\beta}_{\ \ \ \
\ 
\gamma}-Q^{\nu\alpha\beta}_{\ \ \ \ \ 
\gamma}\right)\Delta^\gamma_{\alpha\beta}$$
where the tensor $P_{\mu\nu\rho\sigma}$ is defined in (1) and where
$Q_{\mu\nu\rho\sigma}\equiv2(R_{\rho[\mu}g_{\nu]\sigma}+R_{\sigma[\mu}g_{\nu]\rho})$.
When the vector $\xi^\mu$ is the Killing vector associated
 with time translations with respect to a non rotating frame at
infinity, that is $\xi^\mu=(1,0,0,0,0)$, the associated conserved charge
is the mass
$$M=\int_Sd^{D-2}x\hat J^{[01]}\eqno(13)$$
where $S$ is the $(D-2)$-sphere at infinity. If the vector $\xi^\mu$ is
the
Killing vector associated with rotations with respect to the $\phi$
(resp. $\psi$)
axis, that is $\xi_a^\mu=(0,0,0,1,0)$ (resp. $\xi_b^\mu=(0,0,0,0,1)$),
then the associated charges are the angular momenta.

\vskip 0.8 cm

 To obtain the mass and angular momenta of the 5D-Kerr-like solution of
 EGB theory we inserted the metric (7-9) in the definitions (10-13) and
 took the large $r$ limit  (using {\sl Maple} and {\sl GR tensor}). The result is
$$M_{(5)}=\sqrt{1-{4\tilde\alpha\over
 l^2}}\left[{m\pi\over4(\Xi_a\Xi_b)^2}(2\Xi_a+2\Xi_b-\Xi_a\Xi_b)\right]\ 
 , \  J_{(5)a}=\sqrt{1-{4\tilde\alpha\over
 l^2}}\left[{\pi ma\over2\Xi_a^2\Xi_b}\right]\,.\eqno(14)$$
We calculated similarly the mass and angular momentum in
 $D=6,7,8$ dimensions in the case when there is only one non zero rotation
 parameter. The result is
$$M_D=\sqrt{1-{4\tilde\alpha\over
 l^2}}\, {m {\cal
 V}_{D-2}\over4\pi\Xi_a^2}\left[1+{(D-4)\over2}\,\Xi_a\right]
\quad,\quad
J_{D}=\sqrt{1-{4\tilde\alpha\over
 l^2}}\, {ma {\cal
 V}_{D-2}\over4\pi\Xi_a^2}\eqno(15)$$
where ${\cal V}_{D-2}$ is the volume of the $(D-2)$-unit sphere.
Finally, if one extrapolates to all dimensions the asymptotic
 expressions for the Kerr-AdS Riemann tensor obtained with {\sl Maple}
 and {\sl GR tensor} in $D=5,6,7,8$ dimensions,\footnote{$^{(2)}$}{for
 example : $\lim_{r\to\infty}(R^0_{\ 101}-\overline{R^0_{\
 101}})={m{\cal L}^2\over r^{D+1}}(D-1)[(D-3)W-1]$, where
 $W\equiv\sum_{i=1}^{i=[D/2]}{\mu_i^2\over\Xi_i}$ with
 $\sum_{i=1}^{i=[D/2]}\mu_i^2=1$ and $\Xi_{D/2}=1$ if $D$ is even.}
one finds (by hand) the mass and
 angular momenta of the general D-dimensional Kerr-like solutions of
 Einstein-Gauss-Bonnet theory as
$$
M_{2n}=\sqrt{1-{4\tilde\alpha\over
 l^2}}\left[{m{\cal V}_{2n-2}\over4\pi\Xi}\sum_{i=1}^{i=n-1}{1\over\Xi_i}\right]\quad,\quad
M_{2n+1}=\sqrt{1-{4\tilde\alpha\over
 l^2}}\left[{m{\cal
 V}_{2n-1}\over4\pi\Xi}\left(\sum_{i=1}^{i=n}{1\over\Xi_i}-{1\over2}\right)\right]$$
$$\quad\quad
J_{(D)i}=\sqrt{1-{4\tilde\alpha\over
 l^2}}\left[{m{\cal V}_{D-2}\over4\pi\Xi}{a_i\over\Xi_i}\right]\,.\eqno(16)$$
Note that neither the background nor the vectors
$k^\mu_E$ and $k^\mu_{GB}$ enter the calculation of
 the angular momenta. However they are not given simply by Komar's
 integrals because the first term in (12) is not zero.
Expressions (14-16) have the right limits. When $\alpha=0$ (Einstein's
 theory) they reduce to the results obtained in [2] and [3]. When
 there is no rotation ($a_i=0, \Xi_i=1$) the
masses (16) reduce to the result obtained in [8] and [6] (noting that the
 coefficients of the metric (7) then tend to $g_{tt}\simeq 1/g_{rr}\simeq
r^2/{\cal L}^2-2m/r^{D-3}$).\footnote{$^{(3)}$}{Since the overall factor in (14-16)
 is the
 same as the one which appears in (5) one may conjecture that in the
 general Lovelock theory in $D$ dimensions (whose Lagrangian is the
 sum of the first $[D/2]$
 dimensionally continued Euler forms) the expressions for the masses
 and angular momenta will also be proportional to their values in
 Einstein's theory with an overall coefficient which is zero when the
 equation for the maximally symmetric space curvature has one single
 root of maximal multiplicity $[D/2]$.}

The decisive check of formulae (14-16), that is of the pertinence of
 the proposal (12) made in [6] for the vector $k^\mu_{GB}$, will be
 possible when the full geometry of the Kerr-like rotating {\it black
 holes} is known, by seeing if, with such definitions,  the first law of thermodynamics is still satisfied.

\vskip .5 cm
\noindent {\bf Acknowledgments}
\medskip N.D. thanks Joseph Katz for discussions and the
Yukawa Institute for Theoretical Physics in Kyoto for hospitality and
financial support. Y.M. is supported by a Grant-in-Aid for the 21st
Century COE ``Center for Diversity and Universality in Physics''.

\vskip .5cm

\noindent {\bf References}
\medskip
\item{[1]} G.W. Gibbons, H. Lu, Don N. Page and C.N. Pope, {\it The General Kerr-de
Sitter Metrics in All Dimensions}, hep-th/0404008.
\item{[2]} G.W. Gibbons, M.J. Perry and C.N. Pope, {\it The First Law of
Thermodynamics for Kerr-de Sitter Metrics}, hep-th/0408217.
\item{[3]} N. Deruelle and J.Katz, {\it On the mass of a Kerr-anti-de
Sitter spacetime in D dimensions}, gr-qc/0410135
\item{[4]} J. Katz, {\it A note on Komar's anomalous factor},  Class. Quantum Grav.,
{\bf 2} (1985) 423; J. Katz, J. Bi\v c\'ak, D. Lynden-Bell, {\it Relativistic conservation
laws and integral constraints for large cosmological perturbations}, Phys. Rev. {\bf
D55} (1997) 5759.
\item{[5]} D.G. Boulware and S. Deser, {\it String generated gravity
models}, Phys. Rev. Lett. {\bf 55} (1985) 2656
\item{[6]} N. Deruelle, J. Katz and S. Ogushi, {\it Conserved Charges in Einstein-
Gauss-Bonnet theory}, Class. Quantum Grav., {\bf 21} (2004) 1971,
gr-qc/0310098.
\item{[7]} M.H. Dehghani, {\it Charged Rotating Black Branes in
anti-de Sitter Einstein-Gauss-Bonnet Gravity}, Phys.Rev. {\bf D67} (2003) 064017, hep-th/0211191
\item{[8]}  Rong-Gen Cai,  {\it Gauss-Bonnet Black Holes in AdS
Spaces}, Phys.Rev. {\bf D65} (2002) 084014, hep-th/0109133; I. Neupane, {\it Black hole entropy in string-generated
gravity models}, Phys.Rev.{\bf  D67} (2003) 061501, hep-th/0212092
\end